\documentclass[12pt,epsf]{article}
\usepackage{graphicx}
\usepackage{epsfig}
\begin{document}

\def\a{\alpha}
\def\b{\beta}
\def\c{\varepsilon}
\def\d{\delta}
\def\e{\epsilon}
\def\f{\phi}
\def\g{\gamma}
\def\h{\theta}
\def\k{\kappa}
\def\l{\lambda}
\def\m{\mu}
\def\n{\nu}
\def\p{\psi}
\def\q{\partial}
\def\r{\rho}
\def\s{\sigma}
\def\t{\tau}
\def\u{\upsilon}
\def\v{\varphi}
\def\w{\omega}
\def\x{\xi}
\def\y{\eta}
\def\z{\zeta}
\def\D{\Delta}
\def\G{\Gamma}
\def\H{\Theta}
\def\L{\Lambda}
\def\F{\Phi}
\def\P{\Psi}
\def\S{\Sigma}

\def\o{\over}
\def\beq{\begin{eqnarray}}
\def\eeq{\end{eqnarray}}
\newcommand{\gsim}{ \mathop{}_{\textstyle \sim}^{\textstyle >} }
\newcommand{\lsim}{ \mathop{}_{\textstyle \sim}^{\textstyle <} }
\newcommand{\vev}[1]{ \left\langle {#1} \right\rangle }
\newcommand{\bra}[1]{ \langle {#1} | }
\newcommand{\ket}[1]{ | {#1} \rangle }
\newcommand{\EV}{ {\rm eV} }
\newcommand{\KEV}{ {\rm keV} }
\newcommand{\MEV}{ {\rm MeV} }
\newcommand{\GEV}{ {\rm GeV} }
\newcommand{\TEV}{ {\rm TeV} }
\def\diag{\mathop{\rm diag}\nolimits}
\def\Spin{\mathop{\rm Spin}}
\def\SO{\mathop{\rm SO}}
\def\O{\mathop{\rm O}}
\def\SU{\mathop{\rm SU}}
\def\U{\mathop{\rm U}}
\def\Sp{\mathop{\rm Sp}}
\def\SL{\mathop{\rm SL}}
\def\tr{\mathop{\rm tr}}

\def\IJMP{Int.~J.~Mod.~Phys. }
\def\MPL{Mod.~Phys.~Lett. }
\def\NP{Nucl.~Phys. }
\def\PL{Phys.~Lett. }
\def\PR{Phys.~Rev. }
\def\PRL{Phys.~Rev.~Lett. }
\def\PTP{Prog.~Theor.~Phys. }
\def\ZP{Z.~Phys. }

\newcommand{\bear}{\begin{array}}  
\newcommand {\eear}{\end{array}}
\newcommand{\la}{\left\langle}  
\newcommand{\ra}{\right\rangle}
\newcommand{\non}{\nonumber}  
\newcommand{\ds}{\displaystyle}
\newcommand{\red}{\textcolor{red}}
\newcommand{\mwino}{m_{\widetilde{W}^0}}
\def\ubl{U(1)$_{\rm B-L}$}
\def\REF#1{(\ref{#1})}
\def\lrf#1#2{ \left(\frac{#1}{#2}\right)}
\def\lrfp#1#2#3{ \left(\frac{#1}{#2} \right)^{#3}}
\def\OG#1{ {\cal O}(#1){\rm\,GeV}}


\baselineskip 0.7cm

\begin{titlepage}

\begin{flushright}
IPMU 09-0157
\end{flushright}

\vskip 1.35cm
\begin{center}
{\large \bf
Primordial Black Holes as All Dark Matter
}
\vskip 1.2cm
Paul H. Frampton$^{(a,b)}$, Masahiro Kawasaki$^{(a,c)}$, Fuminobu Takahashi$^{(a)}$, \\and  Tsutomu T. Yanagida$^{(a,d)}$
\vskip 0.4cm

{\it $^{(a)}$ Institute for the Physics and Mathematics of the Universe, 
     University of Tokyo, Chiba 277-8582, Japan\\
$^{(b)}$ Department of Physics and Astronomy, University of North Carolina,
Chapel Hill, NC 27599-3255 \\
$^{(c)}$ Institute for Cosmic Ray Research,
     University of Tokyo, Chiba 277-8582, Japan\\
$^{(d)}$  Department of Physics, University of Tokyo,\\
     Tokyo 113-0033, Japan\\
}

\vskip 1.5cm

\abstract{ 
We argue that a primordial black hole  is a natural and unique
candidate for all dark matter. We show that, in a smooth-hybrid
new double inflation model, a right amount of the primordial black 
holes, with a sharply-defined mass, can be produced at the end
of the smooth-hybrid regime, through preheating. We first consider
masses $< 10^{-7}M_{\odot}$ which are allowed by all the previous 
constraints. We next discuss much heavier mass $10^5 M_{\odot}$ 
hinted at by entropy, and galactic size evolution, arguments. Effects 
on the running of the scalar spectral index are computed.

   }
\end{center}
\end{titlepage}

\setcounter{page}{2}
\section{Introduction}
\label{sec:1}
The presence of dark matter (DM) has been firmly established by a host
of observations, and its abundance was measured by the WMAP with an
unprecedented precision:\cite{WMAP5}
\beq
\Omega_{\rm DM} h^2 \;=\; 0.1131 \pm 0.0034.
\eeq
However it is not known yet what DM is made of, and the question
remains a big mystery in modern cosmology as well as particle physics.

It is often claimed that there is no DM candidate in the framework of
the standard model (SM), assuming that DM is made of elusive particles
which have evaded all conventional DM searches.  The DM particle must
be electrically neutral, long-lived, and cold, but no such particle
exists in the SM.  Thus we need to postulate a theory beyond SM and
introduce a new degree of freedom, which is usually made stable by
imposing an additional discrete symmetry. The introduction of such a
discrete symmetry may be motivated by other phenomenological
reason. For instance, in the supersymmetric standard model (SSM), it
is customary to introduce an R-parity in order to forbid dangerous
operators which would give rise to too fast proton decay. Once the
R-parity is imposed, the lightest supersymmetric particle (LSP)
becomes stable, and therefore a DM candidate.  On the other hand,
there is an argument that the R-parity violation may be a common
phenomenon in the string landscape~\cite{Tatar:2006dc}. If so, the
dangerous operators must be absent due to some other reason(s) and the
lifetime of LSP in the SSM may be too short to account for the DM.

If the DM is made of a weakly interacting massive particle (WIMP), we
may be able to observe collider, direct and indirect DM signatures;
the DM particles may be produced at LHC, and the next-generation
direct search experiments will probe a significant portion of
parameter space predicted by various theoretical DM models.  In spite
of thorough DM searches using widely different techniques, the results
are negative so far.\footnote{One of the exceptions is the DAMA
  experiment~\cite{Bernabei:2000qi}. However, it is still
  controversial concerning the interpretation of the experimental
  data.  Very recently, two DM-like events were found by the CDMS II
  experiment~\cite{Ahmed:2009zw}, but more data is clearly necessary
  to draw definite conclusions.  }  If no DM signature is found in the
future experiments, it may suggest that the basic assumption that the
DM is made of unknown particles is simply wrong.

There actually {\it is} a DM candidate in
the framework of SM, namely, a PBH~\cite{D-objects:PBH}. In the early Universe PBHs can form when
the density perturbation becomes large, and it has been known that a
PBH of mass greater than $10^{15}$\,g survives the Hawking
evaporation~\cite{Hawking:1974sw} and therefore contributes to the DM
density~\cite{Hawking:1971ei}.

In consideration of the entropy of the universe it was pointed out in Ref.~\cite{Frampton:2009nx}
that if all DM were in the form of $10^5 M_{\odot}$ black holes it
would contribute a thousand times more entropy than the supermassive
black holes at galactic centers and hence be a statistically favored
configuration. Here we consider primordial black holes (PBHs) with masses from 
$10^5 M_{\odot}$ to $10^{-8} M_{\odot}$ and, subject to observational 
constraints, any of these masses can comprise all DM although 
the entropy argument favors the heaviest $10^5 M_{\odot}$ mass.

There are several ways to realize large density fluctuations leading
to PBH formation.  One is phase transition involving violent processes
like bubble collision~\cite{Crawford:1982yz,Hawking:1982ga,La:1989st}
or the collapse of string
loops~\cite{Hawking:1987bn,Polnarev:1988dh,Garriga:1993gj,Caldwell:1995fu}.
As we will see in the next section, however, both scenarios have
difficulties.  Another possibility is the production of PBHs from
density fluctuations generated during inflation.  Since the blue
spectrum with a spectral index $n_s > 1$ is disfavored by the WMAP
data~\cite{WMAP5}, a single inflation may not be able to produce large
density fluctuations at small scales unless some dynamics is
introduced during inflation.  On the other hand, the density
fluctuations can be easily enhanced at small scales in a double
inflation model, as first discussed in Ref.~\cite{Kawasaki:1997ju} in
the context of the PBH formation.

In this letter we discuss a double inflation model that consists of a
smooth-hybrid inflation~\cite{Lazarides:1995vr} and a new
inflation~\cite{Izawa:1996dv}. The smooth-hybrid new double inflation
was studied in Ref.~\cite{Yamaguchi:2004tn} in the context of
explaining the large running spectral index suggested by the WMAP 1st
year data~\cite{Spergel:2003cb}.  In this set-up PBHs with a narrow
mass distribution are formed as a result of an explosive particle
production between the two inflations~\cite{Kawasaki:2006zv}. We will show that
the PBH mass can take a wide range of values from $10^{-8} M_\odot$ up
to $10^5 M_\odot$.  Also, the resultant PBH mass has a correlation
with running of spectral index, which was roughly estimated in a semi-analytical method
in Ref.~\cite{Kawaguchi:2007fz}. Here we numerically calculated the correlation, which can be
tested by future observations.

The rest of the letter is organized as follows. In Sec.~\ref{sec:2} we
briefly review the PBH formation and evaporation, and the cosmological
constraints on the PBH abundance.  In Sec.~\ref{sec:3}, we discuss a
realistic double inflation model, namely the smooth-hybrid new
inflation model, and show that PBHs with masses satisfying current
observational constraints are produced. In Sec.~\ref{sec:4} we discuss
related issues and give conclusions.

\section{PBH formation and observational constraints}
\label{sec:2}
The black hole mass and the formation epoch are related to each other
due to the causality.  In the early Universe, the mass contained in
the Hubble horizon sets an upper bound on the PBH mass formed at that
time.  Assuming that the whole mass in the horizon is absorbed into
one black hole, we obtain
\beq
M_{\rm BH} &=&\frac{4 \pi \sqrt{3} M_P^3}{\sqrt{\rho_f}}
 \simeq 0.05\, M_\odot \lrfp{g_*}{100}{-\frac{1}{2}} \lrfp{T_f}{{\rm GeV}}{-2},\non\\
 & \simeq &1.4 \times 10^{13}\, M_\odot   \lrfp{g_*}{100}{-\frac{1}{6}} \lrfp{k_f}{{\rm Mpc}^{-1}}{-2},
 \label{mpbh}
\eeq
where $M_{\rm BH}$ is the black hole mass, $M_P \simeq 2.4 \times
10^{18}$\,GeV is the reduced Planck mass, $M_\odot \simeq 2 \times
10^{33}$\,g is the solar mass, $g_*$ counts the light degrees of
freedom in thermal equilibrium, $\rho_f$, $T_f$ and $k_f$ are the
energy density, the plasma temperature and the comoving wavenumber
corresponding to the Hubble horizon at the formation, respectively.
The radiation domination was assumed in the second equality.

As is well known, Hawking made a striking prediction about the
evaporation of black holes; any black holes have a temperature
inversely proportional to its mass and evaporates in a finite time
$\tau_{\rm BH}$~\cite{Hawking:1974sw},
\beq
\tau_{\rm BH} \;\simeq\;10^{64} \lrfp{M_{\rm BH}}{M_\odot}{3} {\rm yr}.
\eeq
Thus the black holes with mass less than $10^{15}$ g must have
evaporated by now.  PBHs which remain as (a part of) DM must therefore
be created at a temperature below $10^{9}$ GeV. In the following we
assume that PBHs account for all DM in our Universe.

The cosmological effects of PBHs have been extensively studied so
far. While PBHs with masses below $10^{15}$ g are significantly
constrained, it is very difficult to detect PBHs heavier than
$10^{15}$ g because of negligible amount of the Hawking radiation.
The MACHO~\cite{Alcock:1996uj} and EROS~\cite{Renault:1996dp}
collaborations monitored millions of stars in the Magellanic Clouds to
search for microlensing events caused by MAssive Compact Objects
(MACHOs) passing near the line of sight.  The MACHO
collaboration~\cite{Allsman:2000kg} excluded the objects in the mass
range $0.3 M_\odot$ to $30 M_\odot$, and the latest result of the
EROS-1 and EROS-2~\cite{Tisserand:2006zx} excluded the mass range $0.6
\times 10^{-7} M_\odot < M < 15 M_\odot$, as the bulk component of the
galactic DM. On the other hand, if we assume that the PBH formation
occurs before the big bang nucleosynthesis (BBN) epoch, the PBH mass
should be lighter than $10^5 M_{\odot}$ (see Eq.~(\ref{mpbh})).
Therefore we consider PBHs with masses (i) $M_{\rm BH} < 10^{-7}
M_\odot$ and (ii) $30 M_{\odot} <M_{\rm BH} < 10^5 M_\odot$.

Let us comment on other existing constraints.  The PBHs with masses
heavier than $43M_\odot$ were claimed to be excluded by the presence
of wide binaries~\cite{Yoo:2003fr}, but the question on the validity
of the data used to set the limit was raised by
Ref.~\cite{Quinn:2009zg}. Taking account of low averaged DM density
experienced by the four binaries used in their analysis, the strong
constraints set by the wide binaries were undermined.  Recently,
Ricotti, Ostriker and Mack investigated the effect of non-evaporating
PBHs on the cosmic microwave background (CMB) spectrum and anisotropy
and found that the PBHs with mass greater than $\sim 0.1 M_\odot$
cannot account for the bulk component of
DM~\cite{Ricotti:2007au}. However, the authors made assumptions about
accretion efficiency in obtaining strong limits on PBH abundances; if
these assumptions are weakened, all DM could be PBHs for the masses we
consider~\footnote{
As pointed out in Ref.~\cite{Ricotti:2007au}, the Bondi solution becomes invalid
for the PBH of mass $M_{\rm BH} \gsim 10^4 M_\odot$. In particular, the duty cycle 
is not well understood because of the complicated feedback effect.  If the duty cyle is very small, 
the constraint of \cite{Ricotti:2007au} can be weakened. 
}.

The above observational constraints provide us with information on the
PBH formation.  If PBHs are produced at different times, the mass
function tends to be broad, thereby making it difficult to be
consistent with observations. In order to realize the PBH mass
function with a sharp peak, most of the PBHs should be produced at the
same time.  Thus the production mechanism must involve such a dynamics
that only the density fluctuation of a certain wavelength rapidly
grows.

What kind of dynamics can create PBHs?  First of all, density
perturbation must become large for PBHs to be formed.  There are
several ways to realize large density fluctuations leading to the PBH
formation.  One is the phase transition which leads to violent
processes like bubble
collision~\cite{Crawford:1982yz,Hawking:1982ga,La:1989st} or the
collapse of string
loops~\cite{Hawking:1987bn,Polnarev:1988dh,Garriga:1993gj,Caldwell:1995fu}.
However, in the case of the bubble collision, the bubble formation
rate must be tuned to produce the PBH, and the PBH produced from the
strong loops tends to have a broad mass function. Another possibility
is the production of PBHs from density fluctuations generated during
inflation. In the standard picture of inflation, the inflation driven
by a slow-rolling scalar field lasts for more than about $60$
e-foldings to solve theoretical problems of the big bang
cosmology. Then no dynamics for producing a sharp peak in the density
perturbation is expected~\cite{Kawasaki:2004pi}.  However, there is no
a priori reason to believe that our Universe experienced only one
inflationary expansion. Indeed, the cosmological gravitino or modulus
problem can be relaxed if the energy scale of the last inflation is
rather low, and it is then quite likely that there was another
inflation before the last one. If the multiple inflation is a common
phenomenon, we expect that explosive particle production between the
successive inflation periods may produce a sharp peak in the density
perturbation at the desired scales, which leads to the PBH formation
at a later time. In the next section, we show that this is actually
feasible using a concrete double inflation model.

\section{PBHs from preheating}
\label{sec:3}

In this section we provide a double inflation model producing PBHs
with a sharp mass function as an existing proof. The double inflation
model~\cite{Yamaguchi:2004tn,Kawasaki:2006zv,Kawaguchi:2007fz} we
adopt consists of two stages of inflation; the first inflation is
realized by smooth hybrid inflation and the second one by new
inflation. As shown below, the cosmologically relevant density
fluctuations are generated during smooth hybrid inflation.  After the
first inflation, the inflaton and waterfall fields of the smooth
hybrid inflation start to oscillate and decay into their quanta via
self-coupling and mutual coupling of the two fields. The interesting
point is that the decays of the scalar fields are largely enhanced
through parametric resonance and hence the fluctuations of the scalar
fields exponentially grow.  This process is called preheating. During
the preheating phase, only the fluctuations at a specific wavenumber
corresponding to the inflaton mass rapidly grow, and those
fluctuations finally turn into density fluctuations leading to the
production of PBHs with a sharp mass function.  The role of the second
inflation is to stretch the density fluctuations generated during the
first inflation and the subsequent preheating phase to cosmologically
large scales.  In the following, we briefly describe the smooth-hybrid
new inflation model [for details, see
Refs.\cite{Yamaguchi:2004tn,Kawasaki:2006zv}].

The first inflation is realized by smooth hybrid
inflation~\cite{Tetradis:1997kp}.  In
Refs.~\cite{Yamaguchi:2004tn,Kawasaki:2006zv}, the smooth hybrid
inflation model is built in framework of supergravity and the
superpotential and K\"ahler potential are given by
\begin{eqnarray}
  W_{H} & = & S\left(\mu^2 + \frac{(\bar{\Psi}\Psi)^m}{M^{2(m-1)}}\right)
  ~~~~~(m= 2,3,\ldots),\\
  K_H & = & |S|^2 + |\Psi |^2 + |\bar{\Psi}|^2,
\end{eqnarray}
where $S$ is the inflaton superfield, $\Psi$ and $\bar{\Psi}$ are
waterfall superfields, $\mu$ is the inflation scale and $M$ is the
cut-off scale which controls the nonrenormalizable term.  From the
above superpotential and K\"ahler potential together with phase
redefinition and the D-flat condition, we obtain the scalar potential
as
\begin{equation}
   V_H(\sigma, \psi) \;\simeq\;  \left(1+\frac{\sigma^4}{8}+ \frac{\psi^2}{2}\right)
   \left(-\mu^2 + \frac{\psi^{4}}{4M^{2}}\right)^2
   + \frac{\sigma^2\psi^6}{16M^4},
   	\label{eq:pot_hybrid}
\end{equation}
where $\sigma \equiv \sqrt{2}Re S$ and $\psi\equiv 2Re \Psi = 2Re
\bar{\Psi}$.  Here and in what follows we use the Planck unit $M_P=1$
and take $m=2$ for simplicity.  Although the scalar potential
(\ref{eq:pot_hybrid}) is derived in the framework of supergravity, one
may start with (\ref{eq:pot_hybrid}) without assuming supersymmetry.
The potential (\ref{eq:pot_hybrid}) has a true vacuum at $\sigma = 0$
and $\psi = 2\sqrt{\mu M}$. For $\sigma \gsim \sqrt{\mu M}/2$,
however, the potential for $\psi$ has a $\sigma$-dependent minimum at
\begin{equation}
   \psi_{\rm min} \;\simeq\; \frac{2}{\sqrt{3}}\frac{\mu M}{\sigma}.
   	\label{eq:path_hybrid}
\end{equation}
Note that $\psi$ quickly settles down at the minimum during inflation
since its mass is larger than the Hubble parameter.  Then we can
integrated out $\psi$ and obtain the effective potential for $\sigma$
as
\begin{equation}
   V(\sigma) \;=\; \mu^4 \left( 1 + \frac{\sigma^4}{8}
   - \frac{2}{27} \frac{\mu^2 M^2}{\sigma^4}\right)
   = \mu^4+\frac{\mu^4}{8} \left(\sigma^4 - \sigma_d^4
   \left(\frac{\sigma_d}{\sigma}\right)^4\right),
   \label{eq:effective_pot}
\end{equation}
where $\sigma_d \equiv \sqrt{2}/3^{3/8}(\mu M)^{1/4}$.  If the scalar
potential is dominated by the first term, the inflaton $\sigma$ slow
rolls and therefore inflation occurs.

According to the WMAP 5yr data~\cite{WMAP5} , the curvature
perturbation ${\cal R}$, the spectral index $n_s$ and its running
$dn_s/d\ln k$ at the pivot scale $k_* = 0.002 {\rm Mpc}^{-1}$ are
\begin{eqnarray}
   {\cal R} & = & 4.9\times 10^{-5}, 
   	\label{eq:R_obs}\\
   n_s & = & 1.031\pm 0.055, 
   \label{eq:index_obs}\\
   \frac{dn_s}{d\ln k} & = &  -0.037\pm 0.028.
   	\label{eq:running_obs}
\end{eqnarray}
From the effective potential (\ref{eq:effective_pot}) we obtain
\begin{eqnarray}
   {\cal R} & = & \frac{V^{3/2}}{\sqrt{3}\pi V'} =\frac{\mu^2}{\sqrt{3}\pi}
   \left[ \sigma_*^3 + \sigma_d^3\left(\frac{\sigma_d}{\sigma_*}\right)^5
   \right]^{-1}, 
   \label{eq:R_hybrid}\\
   n_s -1 & \simeq  &  2\frac{V''}{V}= \left[ 
   3\sigma_*^2 - 5\sigma_d^2\left(\frac{\sigma_d}{\sigma_*}\right)^6
   \right], 
   \label{eq:index_hybrid}\\
   \frac{dn_s}{d\ln k} & \simeq  & -2\frac{V''' V'}{V^2}= -3\left[ 
   \sigma_*^3 + \sigma_d^3\left(\frac{\sigma_d}{\sigma_*}\right)^5
   \right]
   \left[ 
   \sigma_* + 5\sigma_d\left(\frac{\sigma_d}{\sigma_*}\right)^7
   \right],
   \label{eq:running_hybrid}
\end{eqnarray}
where $\sigma_*$ is the field value of the inflaton
when the fluctuation corresponding to the pivot scale exits the Hubble horizon.

The fluctuation corresponding to the pivot scale $k_*$ exits the horizon at $t=t_*$ when 
$k_*/a(t_*) = H_{H} =\mu^2/\sqrt{3}$ ($H_{H}$: hubble during the smooth hybrid
inflation). Thus the scale factor $a_* = a(t_*)$ is given by
\begin{equation}
    \ln a_* \;=\; -2\ln \mu -136.
\end{equation}
The e-folding number between the horizon exit of the pivot scale
and the end of the smooth hybrid inflation is estimated as
\begin{equation}
  N_*(\sigma) \;=\; \int^{\sigma_*}_{\sigma_e} d\sigma \frac{V}{V'}\simeq 
   \left\{ \begin{array}{ll}
  \ds{   \frac{4}{3\sigma_d^2} -\frac{1}{\sigma_*^2} }& (\sigma_* > \sigma_d)\\[0.8em]
  \ds{   \frac{\sigma_*^6}{3\sigma_d^8} }& ( \sigma_* < \sigma_d)
     \end{array}\right., 
     \label{eq:efold_hybrid}
\end{equation}
where $\sigma_e (\ll \sigma_d)$ denotes the field value when the
smooth hybrid inflation ends.

After the smooth hybrid inflation, $\sigma$ and $\psi$ oscillate about
their minima and decay into the $\sigma$ and $\psi$ quanta via
self-couplings and mutual coupling of the two fields. Since their
effective masses depend on the field amplitudes and therefore
time-dependent, specific modes of the $\sigma$ and $\psi$ quanta are
strongly amplified by parametric resonance. To see this, let us write
down the evolution equation for the Fourier modes of fluctuations
$\sigma_k$ from (\ref{eq:pot_hybrid}) as
\begin{equation}
   \sigma_k'' + 3H \sigma_k' 
   +\left[\frac{k^2}{a^2} + m_{\sigma}^2 
   + 3m_{\sigma}^2\frac{\tilde{\psi}}{\sqrt{\mu M}}
   \cos (m_\sigma t)\right]\sigma_k\; \simeq \;0,
   \label{eq:mathieu}
\end{equation}
where $m_{\sigma} = \sqrt{8\mu^3/M}$ and $\tilde{\psi}$ is the
amplitude of the $\psi$ oscillations. ($\tilde{\psi} \sim \sqrt{\mu
  M}$ at the beginning of the oscillations.)  Neglecting the cosmic
expansion, Eq.~(\ref{eq:mathieu}) has a form similar to the Mathieu
equation which is known to have a exponentially growing solution. The
detailed numerical simulation showed that the wave number for the
fastest growing mode is given by~\cite{Kawasaki:2006zv}
\begin{equation}
  \frac{ k_p}{a_{\rm osc}}\; \simeq\; 0.3 \, m_\sigma.  
   \label{eq:k_peak}
\end{equation}
The width of the peak is also determined by the instability band of Eq.~(\ref{eq:mathieu}),
and it is of $O(0.1) m_\sigma$.

The fluctuations amplified by the parametric resonance eventually
produce PBHs when they reenter the horizon after inflation. The mass
of the PBH is approximately given by the horizon mass when the
fluctuations reenter the horizon. Thus the PBH mass is estimated as
\begin{equation}
  M_{\rm BH}\; \simeq \; 1.4\times 10^{13} \, M_{\odot} \left(\frac{k_p}{{\rm Mpc}^{-1}}
  \right)^{-2}.
  \label{eq:BH_mass}
\end{equation}
From Eqs.~(\ref{eq:k_peak}) and (\ref{eq:BH_mass}) the scale factor at
the beginning of the oscillation phase is estimated as
\begin{equation}
   \ln a_{\rm osc}\; =\; -114 -\ln m_\sigma -0.5 \ln (M_{\rm BH}/M_{\odot}).
\end{equation}
Because the e-folding number $N_*$ is equal to $\ln a_{\rm osc} - \ln a_*$,  we obtain
\begin{equation}
   N_* \;=\; 21 + 0.5 \ln (\mu M) -0.5\ln(M_{\rm BH}/M_{\odot}).
   \label{eq:efold_hybrid2}
\end{equation}

For a fixed black hole mass $M_{\rm BH}$, there are two parameters in
the model, i.e., $\mu$ and $M$, one of which can be removed by using
the WMAP normalization (\ref{eq:R_obs}).  Therefore observable
quantities can be expressed in terms of one free parameter, leading to
a non-trivial relation between $n_s$ and $d n_s/d \ln k$.  In
practice, we adopt $\mu M$ as the free parameter, and solve
Eqs.~(\ref{eq:efold_hybrid}) and (\ref{eq:efold_hybrid2}) for
$\sigma_*$ in terms of $\mu M$. Then $\mu$ and $M$ are determined with
use of Eqs.~(\ref{eq:R_hybrid}) and (\ref{eq:R_obs}) for a fixed $\mu
M$.  Thus, varying $\mu M$, we obtain sets of model parameters which
are consistent with the observed curvature perturbations.

After $\sigma$ and $\psi$ decay, the second inflation ($=$ new
inflation) starts. As mentioned before, the role of the new inflation
is to stretch the fluctuations produced during the smooth hybrid
inflation and subsequent preheating phase to appropriate cosmological
scales. The effective potential for the new inflation is given by
\begin{equation}
  V_{\rm new} \;=\; v^4 \left( 1-\frac{c}{2}\phi^2 \right)-\frac{g}{2}v^2\phi^4
  +\frac{g^2}{16}\phi^8,
\end{equation}
where $\phi$ is the inflaton of the new inflation, $v$ is the scale of
the new inflation and $g$ and $c$ are constants. The scale factor
$a_f$ at the end of the new inflation is estimated as
\begin{equation}
   \ln a_f \;=\; -68 + \frac{1}{3}\ln \left(\frac{T_R}{10^9\GEV}\right)
   -\frac{4}{3}\left(\frac{v}{10^{15}\GEV}\right),
\end{equation}
where $T_R$ is the reheating temperature after the new inflation. 
Therefore, the new inflation should provide the total e-fold number
$\simeq (\ln a_f - \ln a_{\rm osc})$.

\begin{figure}[t]
  \begin{center}
    \includegraphics[width =12cm]{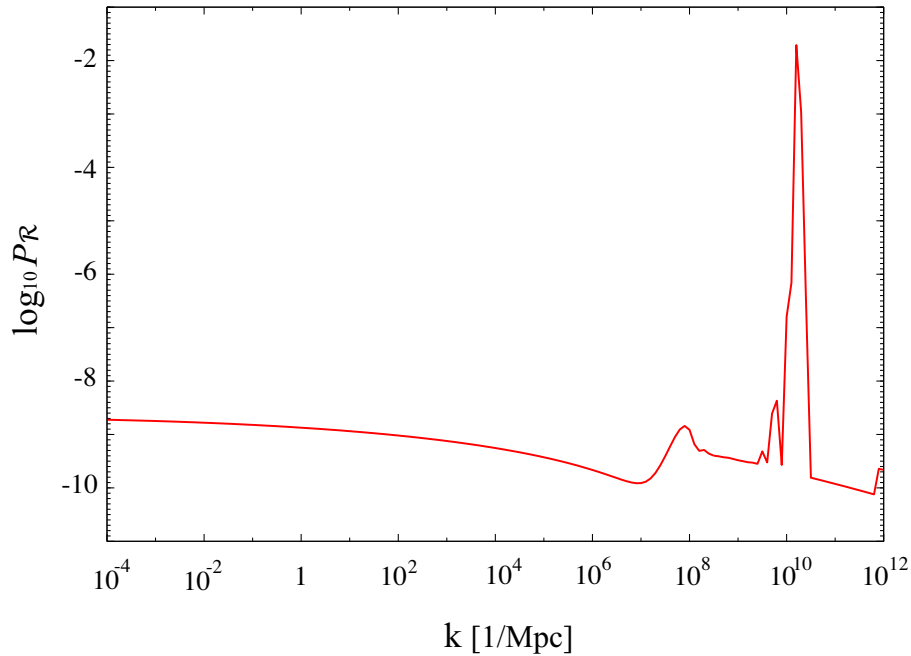}
  \end{center}
  \caption{Power spectrum of the curvature perturbation. We take 
  $\mu= 4 v = 7.14\times 10^{-4}$, $M=0.29$, $c=0.1$ and $g= 2\times 10^{-5}$.}
  \label{fig:power_spec}
\end{figure}

\begin{figure}[t]
  \begin{center}
    \includegraphics[width =12cm]{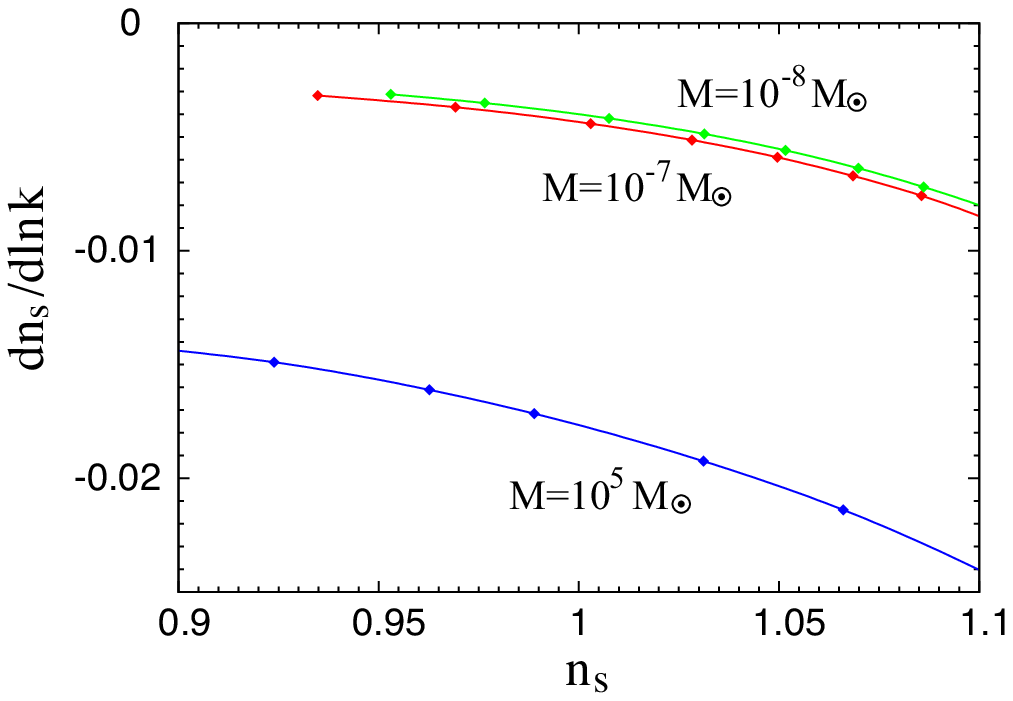}
  \end{center}
  \caption{$n_s$ and $dn_s/d\ln k$ for PBH mass $M_{\rm BH}=10^5 M_\odot, 
  10^{-7}M_{\odot}, $ and $10^{-8}M_{\odot}$. }
  \label{fig:index_running}
\end{figure}

In order to estimate precisely the fluctuations generated in the
present model, we numerically integrate the evolution equations for
the homogeneous modes of $\sigma$, $\psi$, and $\phi$ and the Fourier
modes of their fluctuations as well as the metric perturbations in the
same method described in Ref.~\cite{Kawasaki:2006zv}. In
Fig.~\ref{fig:power_spec} the power spectrum of the curvature
perturbation $P_{\cal R}$ is shown for $\mu= 4 v = 7.14\times
10^{-4}$, $M=0.29$, $c=0.1$ and $g= 2\times 10^{-5}$.  It is seen that
the spectrum has a very sharp peak at $k\simeq 10^{10}\, {\rm
  Mpc}^{-1}$ which corresponds to the PBH mass $\sim
10^{-7}M_{\odot}$. Since the peak is so sharp, PBHs with very
narrow mass range are formed.  The PBH abundance is determined by the height of the peak which depends on the 
decay rate of the inflaton. We can obtain an appropriate PBH abundance  by tuning the
decay rate~\cite{Kawasaki:2006zv}. We also predict the spectral index and
its running for $M_{\rm BH} = 10^5 M_{\odot}, 10^{-7}M_{\odot}$ and
$10^{-8}M_{\odot}$ in Fig.~\ref{fig:index_running}.  Note that the
e-folding number of the smooth hybrid inflation is smaller than that
needed to solve the horizon and flatness problems by a single
inflation model. Therefore the observable density fluctuations are
produced near the end of the inflation when the slow roll parameters
are larger, leading to a larger value of $dn_s/d\ln k$.

\section{Discussion and Conclusions}
\label{sec:4}
In contrast to the conventional WIMP DM model, PBHs have only
gravitational interactions. In order to detect PBHs, we need to
carefully look at the effect induced by PBHs such as gravitational
lensing, gravity waves, etc.  Intermediate mass black holes in the range $30 M_{\odot} < M_{BH} < 10^5 M_{\odot}$
can be sought, for example, by higher-longevity microlensing events~\cite{Alcock:1996uj,Renault:1996dp} and by higher-statistics
analysis of wide binaries~\cite{Yoo:2003fr,Quinn:2009zg}.
In particular there appeared recently an
interesting idea that if the DM is explained by the PBH of mass $10^5
M_\odot$, it may account for the size evolution of the elliptic
galaxies by dynamical friction~\cite{Totani:2009af}.  Further
observations and theoretical study may reveal the presence of the PBH
DM.

What makes the PBH particularly attractive as a DM candidate is that
it is naturally long-lived due to the gravitationally suppressed
evaporation rate. No discrete symmetries need to be introduced in an
ad hoc manner.  Also the PBH DM may be motivated from the arguments
based on entropy of the Universe~\cite{Frampton:2009nx}.

In this letter we have argued that the PBH is a natural and unique
candidate for the DM in the minimal theoretical framework, namely, the
SM. Using the smooth-hybrid new double inflation model, we have shown
that it is possible to produce PBHs of mass ranging from $10^{-8}
M_\odot$ to $10^5 M_\odot$. Importantly, the PBH mass relates the
scalar spectral index and the running of the spectral index, which can
be tested by the Planck satellite.

\section*{Acknowledgment}
MK thanks Tsutomu Takayama for kindly providing his numerical code.
The work of PHF was also supported in part by the U.S.Department of Energy under Grant No.
DE-FG02-06ER41418. The work of  FT was supported by the Grant-in-Aid for Scientific Research on Innovative Areas (No. 21111006) and
JSPS Grant-in-Aid for Young Scientists (B) (No. 21740160).
This work was supported by World Premier International Center Initiative (WPI Program), MEXT, Japan.



\end{document}